# Experimental demonstration of tunable scattering spectra at microwave frequencies in composite media containing CoFeCrSiB glass-coated amorphous ferromagnetic wires and comparison with theory

**D. P. Makhnovskiy and L. V. Panina**

*School of Computing, Communications and Electronics, University of Plymouth,*

*Drake Circus, Plymouth, Devon PL4 8AA, United Kingdom.*

**C. Garcia, A. P. Zhukov, and J. Gonzalez**

*Departamento de Fisica de Materiales, Facultad de Quimica, UPV/EHU, 1072,*

*20080 San Sebastian, Spain.*

**Abstract** We demonstrate composite media with ferromagnetic wires that exhibit a frequency region at the microwave regime with scattering spectra strongly dependent on an external magnetic field or stress. These tunable composite materials have recently been proposed theoretically; however, no direct experimental verification has been reported. We used composite materials with predominantly oriented CoFeCrSiB glass-coated amorphous wires having large magnetoimpedance at GHz frequencies. The free space measurements of reflection and transmission coefficients were conducted in the frequency range 1–8 GHz in the presence of an external static magnetic field or stress applied to the whole sample. In general, the transmission spectra show greater changes in the range of 10dB for a relatively small magnetic field of few Oe or stress of 0.1 MPa. The obtained results are quantitatively consistent with the analytical expressions predicted by the effective medium arguments. The incident electromagnetic wave induces an electrical dipole moment in each wire, the aggregate of which forms the effective dipole response of the whole composite structure in the radiative near or far field region. The field and stress dependences of the effective response arise from a field or tensile stress sensitivity of the ac surface impedance of a ferromagnetic wire. In the vicinity of the antenna resonance the variations in the magneto-impedance of the wire inclusions result in large changes of the total effective response. A number of applications of proposed materials is discussed including the field tunable microwave surfaces and the self-sensing media for the remote non-destructive evaluation of structural materials.

I. Introduction

Composite structural materials containing periodic or random arrays of conducting scattering elements have received much attention because they make it possible to engineer a selective spectral response to the electromagnetic radiation. It has been demonstrated that composites with conducting wires possess a number of unique properties at microwave and optical frequencies, such as anomalous large photonic band gaps,[1] appearance of bulk plasmon modes[2] and resonance-like anomalous dispersion of the effective dielectric function also associated with negative values.[3-6] A recent trend is to achieve the adjustability of these materials. This would be in a great demand in numerous applications, in particular, wireless communication, telemetry and remote non-destructive testing. Several methods were proposed based on biased ferroelectric or ferrite substrates[7] and reconfigurable resonant elements utilizing varactor diodes[8,9] or a system of micro actuators such as the divergent dipoles.[10] These technologies each have its advantages and limitations such as high power consumption, low operational speed, limited frequency band and high cost. It has also been proposed recently that in the composites filled with magnetic wires exhibiting a large magnetoimpedance (MI) effect it should be possible to tune the scattering spectra at GHz frequencies by applying a weak magnetic field or a stress.[11-15] However, this tunability concept has not been experimentally verified. In this paper, using the free space measuring technique in the frequency band of 1–8 GHz, it has been demonstrated that the transmission of the electromagnetic plane wave through a thin sheet of the composite containing CoFe-based amorphous glass-coated microwires can be changed over 15dB by a weak dc magnetic field in the range of 10 Oe or a tensile stress, applied to the whole sample. The experimental results agree well with the theory[11] based on the effective medium concept and the wire electric polarization depending on its magnetic structure.

The main physical principles which determine the behavior of tunable microwave composites filled with ferromagnetic wires have already been established. Wire based composite materials can be treated as continuous media at least in the radiative near or far field region and can be then characterized by the effective permittivity $\varepsilon_{eff}$. If the skin effect in wires is strong, the permittivity $\varepsilon_{eff}$ is determined by the wire geometry, concentration (or lattice period), and the permittivity of the host material. In a general case of a moderate skin effect when the wire radius is in the range of the skin penetration depth $a \approx \delta$, the dispersion characteristics of $\varepsilon_{eff}$ will also depend on the internal electric and magnetic properties of the wires. The influence of the skin effect on the effective dielectric response in diluted wire based materials has been first established theoretically[3,16] and experimentally[4,6] considering nonmagnetic wires with different electric conductivity. It has been then demonstrated theoretically that the controlling quantity is related to the surface impedance of the wires which involves the wire permeability $\mu$ coupled with the dc magnetic structure.[11,12] Therefore, in composite containing ferromagnetic wires exhibiting MI effect at microwave frequencies[17,18] the effective permittivity may depend on a static magnetic field via the corresponding dependence of the surface impedance. The surface impedance can also be changed by applying a stress which modifies the magnetic anisotropy and domain structure in wires. Hence, the effective permittivity $\varepsilon_{eff}$ may also depend on the external stress or strain.

The first experimental attempt to demonstrate the effect of magnetic properties of wires on the permittivity $\varepsilon_{eff}$ was made by measuring reflection/transmission spectra from lattices of long parallel ferromagnetic wires with different magnetic structures whereas the sample topology (the wire length, diameter, spacing in the lattice) was nearly identical.[12] The reported results did demonstrate different permittivity spectra; however, it was not easy to attribute those changes to the wire magnetic structure alone. The experiment was configured to produce the response from "infinitely" long wires since the irradiation was focused in the central portion of the samples. Therefore, the permittivity spectra were expected to be of a plasma type. The composite sample made of wires with an axial magnetization does not show a plasmonic behavior at all, its dielectric spectrum is of a resonance type. The other magnetic composites made of wires with a larger axial anisotropy or a circumferential anisotropy demonstrate a typical plasma-like behavior, however, the plasma frequency which should be mainly determined by the geometrical factor, differs almost twice. Therefore, there is no clear understanding of the effect of magnetic structure on the permittivity spectra.

A natural approach to this problem would be to investigate the possible variations in the scattering response of the same magnetic composite sample under the effect of applied stimuli, such as magnetic field or stress, which could change the magnetization in the wires. However, this direct experiment has not been realized previously because of anticipated problems of applying a magnetic field to a large sample (al least 30cm×30cm) within a free-space installation. We are therefore motivated to experimentally verify the effect of large adjustability of the electromagnetic response of the composites with magnetic wires. The amorphous microwires are soft magnetic materials and their magnetization change can be made with a small magnetic field. For example, in Co-based microwires with a negative magnetostriction and a circular domain structure, the magnetization is rotated towards the axis by a field of a few Oe which is accompanied by a large change in impedance in the range of 100% at the GHz frequencies.[18] Such a field could be easily generated within sufficiently large planar area by a current bus or planar coil wound on the composite sample, placing the current leads perpendicularly to the electrical field vector in the irradiating wave. In our experiments an essentially uniform field up to 12 Oe in the area of 50cm×50cm was produced with a planar coil. The other method to externally control the electromagnetic response from magnetic wire composites is applying a stress or strain. In this case, the negative magnetostrictive wires with a specially induced helical anisotropy should be used. The applied tensile stress would result in changing the anisotropy back to the circumferential direction causing a large decrease in the wire surface impedance.[15] Using a matrix of a high elasticity module, a relatively small stress applied to the whole composite in the range of 0.1 MPa will produce sufficiently large stress acting on the wire via strain transfer.

In this work, we have demonstrated that the composites containing short CoFeCrSiB wires demonstrate a sufficiently narrow bandgap associated with the antenna resonance in wires and negative values of $\varepsilon_{eff}$ beyond the resonance. For circular magnetic configuration in the wire, its magnetic properties play no effect on the propagation of the electromagnetic waves through the composite since the relevant ac permeability is unity. Applying a magnetic field and establishing the magnetization along the axis, corresponds to the case of a high permeability and high surface impedance. This broadens the antenna resonance and opens bandpass: from nearly −27 dB to −12 dB. In the case of composite having wires with induced helical anisotropy, the effect of the tensile stress would be reverse: the application of the stress creates the bandgap.

The proposed technology to control the microwave propagation with magnetic wire structured composite materials is cost effective and is suitable for large scale applications, in particular, as tunable frequency selective surfaces and coatings. The other large area of application includes stress-sensitive materials for microwave non-destructive testing and control in civil engineering. The fiber reinforced materials have become an important class of constructing materials.[19,20] Their performance depends on the bond between the fibers and the matrix, which is not easy to measure or model. Utilizing magnetic wires along with other polymer fibers and measuring microwave response can provide insight on the adhesion quality and stress transfer in the composite, which would be in great demand for the optimization of strengthening characteristics of composites and also for the choice of appropriate matrix.

This paper is structured as follows. In Section II the dispersion characteristics of two types of wire composites are considered demonstrating that the surface impedance of wires can control the scattering spectra. In Section III we review the results on MI in wires relevant to tunable microwave composites. Section IV describes the experimental samples, techniques and results. In Section V we compare the obtained transmission spectra with the theoretical ones and propose future investigations based on a tunable phase of the reflected signal. Finally, in Appendix we demonstrate by rigorous analytical method that the effective permittivity can be applied to wire composites.

**II. Dispersion of the effective permittivity**

In electrodynamics of composites, conductive wire inclusions may induce unusual polarization properties in response to radiation in various frequency ranges over the electromagnetic spectrum. It has been proposed that these structured materials should be treated as continuous media at least in the radiative field region with a characteristic effective permittivity $\varepsilon_{eff}$.[3,4,6] The effective medium approach is also demonstrated in Appendix by a rigorous statistical analysis applied to composites with short parallel wires. The ability of these artificial dielectrics to manipulate the electromagnetic radiation can be then viewed as a consequence of a special dispersion of $\varepsilon_{eff}$.

In natural dielectrics, the electrical polarization is associated with either induced or permanent dipole moments of molecules and atoms. At microwave frequencies the molecular dipole orientation mechanism gives the main contribution to the dispersion of the permittivity which obeys a Debye relaxation. The charge oscillation mechanism and associated resonances become important at much higher frequencies: infrared for the ion polarization and ultraviolet for the electron polarization. The presence of thin conducting wires constrains the electron movement with the associated resonances determined by the geometrical parameters. Then, the frequency dispersion can be established in a specific frequency range from GHz towards visible light. We are interested in composites having the characteristic frequency of a few GHz.

In this work, the main emphasis is placed on composites filled with short conductive wires.[3-6] These materials demonstrate the resonance-like dispersion of the effective permittivity $\varepsilon_{eff}$ with the characteristic frequencies determined by the antenna resonance $f_{res,n} = c(2n-1)/(2l\sqrt{\varepsilon})$, where $c$ is the velocity of light, $l$ is the wire length, $\varepsilon$ is the matrix permittivity (usually, a real constant), and $n \geq 1$ is an integer number (discrete spectrum). In the vicinity of the resonance the real part of $\varepsilon_{eff}$ decreases with increasing the frequency (anomalous dispersion) and may take negative values, whereas the imaginary part has the local maximum. For the wire volume concentrations $p$ below the percolation threshold ($p < p_c \sim 2a/l$, where $a$ is the wire radius),[3] the dispersion properties of these materials can be described in the frame of the Lorentz model for oscillating electrical dipole moments with a damping. In our case, the wire inclusions play the role of "elementary scatterers" having a dipole moment induced by the external electromagnetic wave. The general form of the permittivity is:

$$\varepsilon_{eff}(\omega) = \varepsilon + 4\pi\, p \sum_{n=1}^{\infty} \frac{A_n}{(\omega_{res,n}^2 - \omega^2) - i\omega\Gamma_n}, \qquad (1)$$

where the summation is carried out over all the antenna resonance frequencies $\omega_{res,n} = 2\pi f_{res,n}$, $i = \sqrt{-1}$ is the complex unit, the strength of oscillators and relaxation are described by the phenomenological parameters $A_n$ and $\Gamma_n$, respectively.[21,22] The main relaxation channels in the system are resistive and magnetic losses, and also the wave scattering.[11] Each phenomenological parameter $\Gamma_n$ includes all of these relaxation contributions. The permittivity $\varepsilon_{eff}$ by itself is not a measurable quantity in the wave scattering experiments. However, it can be deduced from the measured complex-valued $S$-parameters. As it has been demonstrated in experimental works Ref. 4 and Ref. 6, $\varepsilon_{eff}$ recovered in such manner is well approximated by the Lorentz dispersion characteristic (1), where only the first term is important, since $A_1 \gg A_{n>1}$. The resonance contribution to $\varepsilon_{eff}$ can be very large even for small concentrations $p < 0.01\%$. Then, $\varepsilon_{eff}$ becomes negative in some frequency band past the resonance and the wave propagation is forbidden in this frequency range. If the relaxation in the system is large, the dispersion of $\varepsilon_{eff}$ broadens and its real part stays positive. In this case, the bandgap will be essentially eliminated. As it has been demonstrated in Ref. 11, in the case of a moderate skin effect the internal losses (resistive and magnetic) are dominant, whereas the radiation losses can be neglected. The magnetic contribution to $\Gamma_n$ occurs because of the following. When the composite is irradiated by the electrical field $e_0$ parallel to the wires, this field also creates a circular magnetic field $h_\varphi$ within the wire (the wire axis is denoted as $x$-axis). At the wire surface, the fields $e_0$ and $h_\varphi$ are related via the longitudinal component $\varsigma_{xx}$ of the total wire impedance tensor:[23]

$$e_0 = \varsigma_{xx} h_\varphi. \qquad (2)$$

Equation (2) should be taken as a boundary condition when solving the scattering problem from a conducting wire. In a magnetic wire, the parameter $\varsigma_{xx}$ depends on the permeability as a response to $h_\varphi$. The circular permeability has a maximum when the magnetization in the wires is along its axis. On the contrary, the field $h_\varphi$ does not influence a circular magnetization (the domain wall movements at GHz frequencies should be neglected) and the impedance does not depend on the magnetic properties. The first case corresponds to high impedance and strongly increased damping resulting in very broad dipole resonances. What makes the amorphous microwires special is that the transition from a circular magnetization to an axial magnetization (and hence, the state of low and high impedances) can be realized by applying a very small magnetic field, in the order of just few Oe. The other aspect which is important for tunability is that due to amorphous state of the wire, its magnetic anisotropy is mainly determined by the magnetostrictive interaction. This makes it possible to establish a certain magnetic anisotropy by proper annealing (or thermal) treatment and change it by applying an external stress. The dependence of the surface impedance on the magnetic properties of the wire will be considered in the following section. If the interaction between the short wire inclusions is neglected, the average polarization $\eta$ of a wire and the effective permittivity $\varepsilon_{eff}$ of the whole composite take simple analytical forms:[14]

$$\eta = \frac{1}{2\pi \ln(l/a)(\tilde{k}a)^2}\left(\frac{2}{\tilde{k}l}\tan(\tilde{k}l/2) - 1\right), \quad (3)$$

$$\varepsilon_{eff} = \varepsilon + 4\pi\, p\eta\, .$$

In Eq. (3), the susceptibility was derived in the assumption of that the magnetic and resistive losses prevail over the radiation ones. Here $a$ is the wire radius and $\tilde{k}$ is the renormalized wave number depending on the surface impedance in the following way:

$$\tilde{k} = k\left(1 + \frac{ic\varsigma_{xx}}{\omega\, a \ln(l/a)}\right)^{1/2},$$

where $k = \omega\sqrt{\varepsilon}/c$ is the wave number of the dielectric matrix. It could be shown that this renormalization is important only in the case of a moderate skin effect.

For comparison, we now consider the effective permittivity in the system of long (continuous) wires. It has a principle difference when compared with the Lorentz model, in that the motion of free electrons is not restricted in the longitudinal direction and the composite with continuous wires can be considered as "the cold plasma". On the contrary, in composite with short wires the free electrons can move only within the wire length having their maximum density at the wire ends. This "longitudinal localization" was responsible for the dipole moment defining the polarization of the wire inclusions in Eq. (3). However, confining the electrons within thin wires has two consequences: reduced electron density and increased effective mass. As a result, the effective permittivity for waves with the electric field polarization along the wires has a characteristic plasma dispersion behavior but with strongly reduced plasma frequency $\omega_p$:[2]

$$\varepsilon_{eff\|} = \varepsilon - \frac{\omega_p^2}{\omega^2}, \tag{4}$$

$$\omega_p = \frac{2\pi c^2}{b^2 \ln\left(\frac{b}{a}\right)}, \tag{5}$$

where $b$ is the cell parameter (the average distance between the wires) and $\varepsilon_{eff\|}$ is the effective permittivity when the wave electrical field is directed along the wires. Equation (4) entirely corresponds to the model that describes the dielectric response from the plasma of free electrons. The value of the plasma frequency would be in the GHz range for the radius $a$ of few microns and the cell parameter $b$ of few millimeters. For $\omega < \omega_p/\sqrt{\varepsilon}$, $\varepsilon_{eff\|}$ becomes negative.

The effective permittivity of the system of parallel wires could be also obtained from the first principles by solving the Maxwell equations and employing a homogenization procedure.[16] This approach was applied to the system of non-magnetic conducting wires and takes account of the field distribution inside them:

$$\varepsilon_{\it{eff}\parallel} = \varepsilon - p\frac{2\varepsilon_c F_1(k_c a)}{(ak_c)^2 F_1(k_c a)\ln\left(\frac{b}{a}\right) - 1}, \tag{6}$$

$$F_1 = J_1(x)/xJ_0(x),$$

where $p = \pi a^2/b^2$ is the wire volume concentration, $J_{0,1}(x)$ are the Bessel functions, $k_c^2 = 4\pi i\omega\sigma/c^2$ is the wave number in a conducting wire, $\sigma$ is the wire conductivity, and $\varepsilon_c = 4\pi i\sigma/\omega$ is the dielectric permittivity of the conductor. In the case of a strong skin effect ($ak_c \sim a/\delta \gg 1$) equation (6) reduces to (5) and becomes independent of the internal parameters of the wires. The approach developed in Ref. 16 can be extended for the case of magnetic wires. It is possible to write the solution of the electric field in the system which depends on the wire property using the surface impedance which would contain all the information on the conducting and magnetic properties of the wire. In this way, Eq. (6) can be represented as following:

$$\varepsilon_{\it{eff}\parallel} = \varepsilon - p\frac{\omega_p^2}{\omega^2\left(1 + i\frac{c\varsigma_{xx}}{\omega a}\ln\left(\frac{b}{a}\right)\right)}. \tag{7}$$

Therefore, we have demonstrated that in both types of wire-based composites, their effective permittivity shows unusual behavior in the GHz frequency range which can be controlled by changing the surface impedance in the wires. In Refs. 11 and 12 it was proposed for the first time to use the field dependent impedance in the ferromagnetic wire with a specific magnetic anisotropy to change the microwave response.

## III. Magnetoimpedance (MI) in amorphous magnetostrictive wires

We have demonstrated that the microwave dielectric response of structured composites may depend on magnetic properties of constituent elements via the surface impedance, which can be used as the boundary condition (2) to find the electromagnetic field distribution outside the filling inclusions. In the case of a strong skin effect in the inclusions and a scalar permeability $\mu$ the surface impedance $\varsigma$ is determined by the Leontovich condition:

$$\varsigma_{xx} = (1-i)\sqrt{\frac{\omega\mu}{8\pi\sigma}}. \tag{8}$$

In fact, Eq. (8) has an illustrative character. First of all, the condition of a strong skin effect may not be valid or may not represent the case of physical interest. In our case, when the skin effect becomes strong the effective permittivity will mainly depend on the structural parameters of the system and not on the internal properties of the inclusions. The other difficulty is related with the permeability calculations. This parameter depends on the dynamic magnetization processes involved which could be quite complicated in real systems resulting in a particular frequency and spatial dispersion of $\mu$. Furthermore, the permeability and hence impedance have a tensor form even in uniform materials because of the magnetization precession. The detailed analysis of the surface impedance in magnetic wires with uniform magnetization (in a single domain state) valid for any frequency and type of magnetic anisotropy has been made in Refs. 23 and 24. The calculation of the impedance tensor $\hat{\varsigma}$ was based on the solution of the Maxwell equations for the ac electric **e** and magnetic **h** fields inside the wire together with the equation of motion for the magnetization vector **M**. Considering a linear approximation with respect to the time-variable parameters **e**, **h**, $\mathbf{m} = \mathbf{M} - \mathbf{M}_0$, where $\mathbf{M}_0$ is the static magnetization, and neglecting the exchange effects, the problem was simplified to finding the solutions of the Maxwell equations with a given ac permeability tensor $\hat{\mu} = 1 + 4\pi\hat{\chi}$, $\mathbf{m} = \hat{\chi}\mathbf{h}$. In a single domain wire the linear magnetization dynamics is described by a uniform precession of the total magnetization vector $\mathbf{M}$ around $\mathbf{M}_0$. The susceptibility tensor is obtained from the linearized Landau-Lifshitz equation considering the magnetic potential energy of the form:

$$U_m = -K_{eff}(\mathbf{n}_K \cdot \mathbf{M})^2 / M_0^2 - (\mathbf{H} \cdot \mathbf{M}), \tag{9}$$

$$\mathbf{H} = \mathbf{H}_{ex} + \mathbf{h},$$

where $(\mathbf{n}_K \cdot \mathbf{M})$ and $(\mathbf{H} \cdot \mathbf{M})$ are the scalar products. In Eq. (9) a uniaxial magnetic anisotropy is considered with the effective anisotropy constant $K_{eff}$ and easy axis $\mathbf{n}_K$ having some angle $\alpha$ with the wire $x$-axis. The external magnetic field $\mathbf{H}$ includes the dc field $\mathbf{H}_{ex}$ (which will be used as an external factor to change the impedance and microwave response) and the high frequency field $\mathbf{h}$. Because of the demagnetizing effects, only the projection of $\mathbf{H}_{ex}$ on the wire axis will influence the magnetic configuration. In amorphous materials the magnetic anisotropy is mainly determined by the magnetoelastic interaction and will depend on the external stress or strain that also can be use as a tuning parameter. In the coordinate system with the axis $x'$ directed along the static magnetization, the susceptibility tensor has the following form with respect to the unit vector $(n_z, n_y, n_x)$:

$$\hat{\chi} = \begin{pmatrix} \chi_1 & -i\chi_a & 0 \\ i\chi_a & \chi_2 & 0 \\ 0 & 0 & 0 \end{pmatrix}. \tag{10}$$

The susceptibility components $\chi_1$, $\chi_2$, and $\chi_a$ depend on the frequency $\omega$ and the magnetic configuration as following:

$$\chi_1 = \omega_M(\omega_1 - i\tau\omega)/\Delta,$$

$$\chi_2 = \omega_M (\omega_2 - i\tau\omega)/\Delta,$$

$$\chi_a = \omega\, \omega_M /\Delta,$$

$$\Delta = (\omega_2 - i\tau\omega)(\omega_1 - i\tau\omega) - \omega^2, \tag{11}$$

$$\omega_1 = \gamma\left(H_{ex}\cos(\theta) + H_K \cos(2(\alpha - \theta))\right),$$

$$\omega_2 = \gamma\left(H_{ex}\cos(\theta) + H_K \cos^2(\alpha - \theta)\right),$$

$$\omega_M = \gamma M_0, \quad H_K = 2K_{eff}/M_0.$$

Here, $\gamma$ is the gyromagnetic constant, $\tau$ is the spin-relaxation parameter, $H_K = 2K_{eff}/M_0$ is the anisotropy field, $H_{ex}$ is the external magnetic field along the wire axis, and $\theta$ is the angle between the dc magnetization $\mathbf{M}_0$ and the wire axis. Solving the Maxwell equations with permeability tensor (10) allows the surface impedance to be expressed as the asymptotic series with respect to the parameter $a/\delta$, where $\delta = c/\sqrt{2\pi\sigma\omega}$ is the non magnetic penetration depth. In the case of a strong skin effect ($a/\delta \gg 1$) the longitudinal component of the impedance takes the following form:

$$\varsigma_{xx} = (1-i)\sqrt{\frac{\omega}{8\pi\sigma}}\left(\sqrt{\tilde{\mu}}\cos^2(\theta) + \sin^2(\theta) + \frac{\delta(1+i)}{4a}\right), \tag{12}$$

$$\tilde{\chi} = \chi_2 - \frac{4\pi\,\chi_a^2}{1 + 4\pi\,\chi_1}, \tag{13}$$

$$\tilde{\mu} = 1 + 4\pi\,\tilde{\chi},$$

where $\tilde{\chi}$ and $\tilde{\mu}$ are the effective susceptibility and permeability, respectively, as a response to circular excitation field $h'_\varphi$ in the prime coordinate system ($x' \| \mathbf{M}_0$). The first two terms of equation (12) correspond to Eq. (8) where $\sqrt{\mu}$ should be replaced with $\sqrt{\tilde{\mu}}\cos^2\theta + \sin^2\theta$. Therefore, the high frequency impedance in a magnetic wire depends on both the ac permeability and static magnetization angle. However, for low magnetic fields, with increasing frequency the permeability tends to be unity and the dependence on $\theta$ vanishes. The last term in Eq. (12) gives evidence that the actual expansion parameter is $\delta_m/a$ with a magnetic skin depth determined as $\delta_m = \delta/\sqrt{\tilde{\mu}}$. It means that for $\tilde{\mu} >> 1$ Eq. (8) is valid for much lower frequencies than that determined by the condition $\delta/a << 1$. In the case of amorphous wires with the conductivity of $\sigma = 10^{16} s^{-1}$, the non-magnetic skin depth is about 10 microns at 2 GHz, whereas $\delta_m$ calculated with the value of $\tilde{\mu}$ defined from Eq. (13) with parameters typical of amorphous wires is 1.2 microns. Since the permeability value decreases with frequency the value of $\delta_m$ changes little in the frequency range $1-10$ GHz. This estimation shows that the high frequency asymptote may be not satisfactory even in this frequency range for very thin wires with the radius of few microns. In this case, the asymptotic solution for the opposite limit ($a/\delta << 1$) should be used:

$$\varsigma_{xx} = \frac{k_m c}{4\pi\sigma} \frac{J_0(k_m a)}{J_1(k_m a)} + \frac{1}{54}\left(\frac{a}{\delta}\right)^4 \frac{c\mu_3^2}{\pi\sigma\, a}. \tag{14}$$

Here for the reader convenience the designations of Ref. 23 are used: $k_m^2 = \mu_1\left(4\pi i \omega \sigma/c^2\right)$, $\mu_1 = 1 + 4\pi \cos^2(\theta)\tilde{\chi}$, $\mu_3 = -4\pi \sin(\theta)\cos(\theta)\tilde{\chi}$.

In Refs. 11 and 18 it was shown that in the microwave range a weak external magnetic field $H_{ex} < H_K \leq 10$ Oe or a moderate stress of less than 1 GPa does not result in noticeable changes in the permeability $\tilde{\mu}$. Therefore, the change in high frequency impedance can be realized by controlling the angle $\theta$ of the static magnetization $\mathbf{M}_0$ with respect to the wire $x$-axis. Because of this, unusual property of the high frequency magnetoimpedance was referred to as a "valve-impedance" $\varsigma_{xx} \sim \cos^2(\theta)$. In the MHz range this dependence becomes more complicated because $\tilde{\chi}$ also depends on the external magnetic field or stress.

Amorphous magnetic wires of Co-rich compositions having a small negative magnetostriction $\lambda \approx -10^{-7}$ exhibit the valve-impedance in the range of 100% at 1–2 GHz and they would be most suitable to realize magnetically tunable microwave composites. For a general analysis of the dc magnetization process under the effect of the external field $H_{ex}$ and/or a tensile stress $\sigma_{ex}$ we assume that the effective anisotropy is contributed by some axial anisotropy $K$, an axial stress $\sigma_a = \sigma_{res} + \sigma_{ex}$ which includes the residual $\sigma_{res}$ and external $\sigma_{ex}$ components, and also the torsion $\sigma_t$. All theses contributions to the anisotropy are considered to be uniform inside the wire. This is certainly a simplified model; nevertheless, it describes quantitatively well the experimental results on hysteresis loops and impedance. The residual stress and torsion are typically introduced by the wire drawing process with $\sigma_{res} \gg \sigma_t$. The latter, however, should not be neglected. Because of torsion, the magnetization is deviated from the ultimate states: axial or circumferential and the hysteresis process is described by a smooth function. The torsion can be partitioned into two perpendicular stresses $\pm \sigma_t$, each acting at 45° relative to the wire axis. The positive $\sigma_t$ is the tensile stress, whereas the negative $-\sigma_t$ is the compressive stress. Then, Eq. (9) for the magnetostatic energy $U_m$ is written in the following form:[15]

$$U_m = -K\cos^2\theta - \frac{3}{2}\lambda\sigma_a \cos^2\theta - \frac{3}{2}\lambda\sigma_t \left(\cos^2(\theta-45°) - \cos^2(\theta+45°)\right) - M_0 H_{ex}\cos\theta .$$

After some algebra, this equation can be converted into the equivalent uniaxial anisotropy form with the anisotropy angle $\alpha$ measured from the wire $x$-axis:

$$U_m = -\left|\widetilde{K}\right|\cos^2(\alpha-\theta) - M_0 H_{ex}\cos\theta . \qquad (15)$$

Here:

$$\widetilde{K} = \frac{K + (3/2)\lambda\sigma_a}{\cos(2\widetilde{\alpha})},$$

$$\tilde{\alpha} = \frac{1}{2}\tan^{-1}\frac{3|\lambda\sigma_t|}{|K+(3/2)\lambda\sigma_a|}.$$

At that, the anisotropy angle $\alpha$ is chosen as following:

a) if $K+(3/2)\lambda\sigma_a > 0$, then $\alpha = \tilde{\alpha}$,

b) if $K+(3/2)\lambda\sigma_a = 0$, then $\alpha = 45°$ and $\tilde{K} = 3\lambda\sigma_t$

c) if $K+(3/2)\lambda\sigma_a < 0$, then $\alpha = 90° - \tilde{\alpha}$.

For a negative magnetostriction ($\lambda < 0$), small values of $K$ and $\sigma_t$ ($K < (3/2)|\lambda|\sigma_{res}$, $\sigma_t << \sigma_{res}$) the anisotropy is nearly circumferential. Applying an axial magnetic field larger than the anisotropy field $H_K$ (which is of the order of few Oe in some CoFe-based amorphous wires) establishes the axial magnetization, thus, enabling to change the wire microwave impedance.[17,18] For such anisotropy, however, the application of a tensile stress (when $H_{ex} = 0$) does not change the magnetization and hence its effect on the impedance is small. In the opposite limit when the uniaxial anisotropy constant $K$ is large, the situation is reverse: the overall anisotropy is nearly axial anisotropy, the effect of $H_{ex}$ (when $\sigma_{ex} = 0$) is small and the tuning parameter becomes $\sigma_{ex}$. The application of the tensile stress will rotate the anisotropy axis and the magnetization towards the circumference, that will result in the dependence of $Z$ on $\sigma_{ex}$ in zero field.[15] In the intermediate case of a helical anisotropy, both $H_{ex}$ and $\sigma_{ex}$ would cause a change in the static magnetization but the sensitivity of the impedance change would be smaller. It should be noted that the establishment of the axial anisotropy in negative magnetostrictive wires or a circumferential anisotropy in positive magnetostrictive wires is a complicated process which requires stress annealing at a certain temperature regime and cooling under stress.[17,25] It was assumed that during this process the resulting residual stress becomes compressive although annealing was performed under tension. The establishment of such "reverse anisotropy" during this process was confirmed by measurement of the hysteresis loops and impedance. It has also been reported recently the existence of a dominant axial anisotropy in as-prepared negative magnetostrictive CoFeSiB wires having relatively thick glass coating of 10 microns which is nearly equal to the metal core radius.[26] Therefore, magnetic anisotropy in amorphous glass coated wires could be tailored to get a preferable magnetic structure. The exact mechanism of forming a "reverse" anisotropy is not clear and could be attributed to a creep induced anisotropy or compressive stress.

## IV. Experimental

### IV.I Wires and their surface magnetoimpedance

Composite materials demonstrating tunable microwave response were fabricated using $Co_{68}Fe_4Cr_3B_{14}Si_{11}$ glass coated amorphous wires. This alloy has a small negative magnetostriction of about $-10^{-7}$. As-prepared wires typically have a nearly circumferential anisotropy due to a predominant axial residual stress coupled with a negative magnetostriction ($K \approx 0$, $\sigma_t \approx 0$). The circumferential anisotropy becomes better established (smaller anisotropy deviation from circumference) by heating wires under a small tension due to stress relaxation. In our case the wire was mounted on an aluminium spool under tension of 20-30 kg/cm$^2$ and the whole spool was heated at a temperature of 160–250$^0$ C for about 10–30 min. These wires will be suitable to realize composite materials demonstrating field dependent microwave response. The metal core diameter was 22 μm and the total diameter was 35 μm. For stress sensitive composite materials, it is important to have a substantial deviation of the anisotropy from the circumferential direction (in the ideal case axial anisotropy would be preferred). The type of the anisotropy was modified by heating the wire under a larger applied stress with the following slow cooling also under tension. For this treatment, the wires of the same composition as above were used but with a smaller total diameter of 20.3 μm and the metal core diameter of 14.5 μm. The wires were heated at a temperature of $250^0$ C for 30 min under a tension of 50–70 kg/cm$^2$. This thermo mechanical treatment creates a helical anisotropy.

The impedance $Z = \varsigma_{xx}(2l/ca)$ of a short wire sample of 6 mm in length was obtained from $S_{11}$ parameter (reflection coefficient) measured by means of a HP8720ES Vector Network Analyzer with a coaxial sample holder in which the wire is placed as a central conductor. The measurements were made for frequencies from 10 MHz to 6 GHz, under different tensile stress and external field in the range $\pm 50$ Oe supplied by a solenoid parallel to the wire axis. One end of the wire is soldered to a SMA connector and the other to a non-magnetic movable bar with a micrometer screw to apply a tensile stress (up to 900 MPa). To calculate the applied stress, the Young modulus of the composite structure (amorphous metal core coated with glass) was measured in an Instron 4400 Tensile/Compression Test System. For the wire geometry used the Young modulus was found to be 170 GPa.

The impedance plots vs. $H_{ex}$ are shown in Figures 1(a) and 1(b) for two frequencies of 50 MHz and 2 GHz, respectively. The impedance data are given for the stress-annealed wires which exhibit a substantial dependence on the applied tensional stress. In both cases of a relatively low frequency of 50 MHz and high frequency of 2 GHz the application of the stress of 900 MPa substantially lowers (by 45% and 36%, respectively) the value of the impedance at zero field. It also can be seen that the stress sensitivity of the impedance increases in the presence of a small magnetic field. At lower frequency of 50 MHz, the field behavior of the impedance is characterized by a minimum at zero field and two symmetrical peaks seen at the field nearly equal to the anisotropy field of 2.4 Oe for $\sigma_{ex}=0$ and 5 Oe for $\sigma_{ex}=900$ MPa. When the frequency is increased towards GHz frequency range, the impedance behavior for fields smaller than the anisotropy field changes little, but the peaks flatten and for a frequency of 2 GHz there is a very slow increase in impedance for higher fields. The impedance plots vs. field seem to be typical of wires with a circumferential anisotropy or a helical anisotropy as discussed above in Section III. The decrease in impedance under the effect of the applied tension can be related with decrease in the value of $\cos\theta$ (see Eq. (12)) meaning that the angle between the static magnetization and the wire axis increases, thus this indicates that a helical anisotropy exists in stress annealed wires which transforms to a circumferential one in the presence of $\sigma_{ex}$. In the case of thicker wires annealed with a small stress, the minimum in impedance at zero field is much deeper and the applied stress has no effect on its value. In this case the annealing treatment helps to refine the circumferential anisotropy and the relative change in impedance is as much as 1000% per 5 Oe at frequency of 50 MHz (data are not shown).

**IV.II Composite manufacture**

The wires described above were used to fabricate two types of composite materials to demonstrate the effect of external stress and/or magnetic field applied to a whole material on microwave response. The wire pieces of 5 cm in length were situated between two sheets of elastic fabric with in plane size of 34cm×38cm. One layer of fabric was soaked in silicon glue, while the wire inclusions were distributed on the second dry layer. The wire pieces were predominantly aligned in one direction as shown in Fig. 2. The number of wire centers (a point at half length) per cm$^2$ is convenient to use as a characteristic planar concentration $\bar{p}$ of the inclusions. In the tested composite this concentration was kept very small, not exceeding 0.5 centers per square centimeter. The wires were covered with the soaked layer to form a sandwiched structure which was then placed under a gentle press until all the glue was cured. After polymerization the structure preserves its flexibility. The total thickness of the composite sample was about 2 mm. The sample was attached to two strips of wood, which allows the uniform stress application with a load $P$.

It seems that the application of the external magnetic field to such a large sample would be impractical and this is why in Ref. 12 the magnetic tunability was demonstrated indirectly by using wires with different dc magnetic structure. In this work, a magnetic field as large as 12 Oe was applied to the whole composite using a specially designed coil. We have used a planar coil of size 50cm×50 cm with the wire diameter of 0.8 mm and turns mount with a relatively large step $d$ of 1 cm (50 turns). To provide the maximum uniformity of the generated field, this distance should be equal to that between the coil layers (as for the Helmholtz coil). Yet, the field alteration along the coil length remains very large. To avoid this, the coil turns in the two layers have to be shifted with respect to each other at $d/2$, as shown in Fig. 3. If the wire inclusions are ordered in one direction, the coil turns have to be perpendicular to the wires to generate the field along them. For the wire inclusions with randomly distributed angles the "valve-like" behavior of the magnetoimpedance $\varsigma_{xx} \sim \cos^2(\theta)$ will guarantee that for sufficiently larger magnetic field $H_{ex} > H_K$ most of wires will be switched to the high impedance state. In addition, the coil turns have to be perpendicular to the electric field of the incident radiating wave to ensure the wave transmission through. It should be noticed that such a coil can be easily integrated with the composite sample to constitute a single structural element. For a coil wire diameter of 1 mm and a current of 10 A, the 50 cm coil consumes 106 W and induces a sufficiently uniform magnetic field of 12 Oe. The other way to apply the field to a large planar sample is with the help of a current bus. However, since the field would be generated in large volume (not concentrated in the surface layer) its value is small for realistic configurations and power consumption being in the range of 1 Oe, which is less than the anisotropy field in the wires.

**IV.III Free-space measurement technique and transmission spectra**

The measurements of the effective microwave properties of composites were conducted in free space using standard technique of the TRL calibration (Trough-Reflection-Line), which has been successfully explored for a long time in testing dielectric materials.[27-31] The measuring system scheme is shown in Fig. 4. The system includes a pair of broad band antennas connected to the Analyzer's ports and a mini anechoic chamber covered inside with a microwave absorber. In general, it is possible to avoid the parasitic scattering in the measuring track without use of an anechoic chamber by employing a time domain option,[27] which enables the separation of the useful signal from background scattering. The broad band horn antennas have the following characteristics: length 887 mm, aperture 351mm×265mm, a frequency range of $0.85-17.44$ GHz, SWR<1.7, and the effective area more than 150 $mm^2$ in the range of 0.85–15 GHz. The front walls of the chamber, on which the antennas were fastened, were made mobile to enable the distance between the antennas to be variable. This is necessary for the preliminary TRL calibration of the measuring track. A thin planar composite sample is fastened on fiberglass ropes in the middle of chamber at a distance of 40 cm from each horn antenna. This space guaranties the measurements in the radiative near field for the frequencies higher than 1 GHz. In our experiments we do not use focusing lenses to obtain the averaged response from many wire inclusions.

For the study of tunable properties, two types of external stimuli were applied to the composite sample: (i) dc magnetic field $H_{ex}$ induced by a planar coil and (ii) tensile stress $\sigma_{com}$ applied to the sample frame by means of a load hanged outside the chamber. The maximal value of $H_{ex}$ was 12 Oe and the maximal value of $\sigma_{com}$ was 0.1 MPa. Since silicone materials have a relatively low stiffness of $50-100$ MPa the stress transferred through the matrix strain to the wires was very large, in the range of GPa, which would be sufficient to saturate the magnetization in wires with a helical anisotropy in the circumferential direction, as discussed above (see Fig. 1). A precise calculation of the stress imposed on wires would require special investigation to take account of the effect of wires on the matrix stiffness and adhesion between the wires and matrix. The transmission/reflection spectra are then measured and plotted for different values of $H_{ex}$ and $\sigma_{com}$. In this work, we were not in a position to deduce the effective permittivity parameter from these spectra. However, in the next Section we will make comparison of the experimental results with the theoretical ones obtained on the basis of the effective permittivity model.

The spectra of the transmission coefficient ($S_{21}$ or $S_{12}$) through the composite sample containing wires with a circumferential anisotropy are shown in Fig. 5 for two values of the planar concentration $\bar{p}=0.35$ and $\bar{p}=0.46$ centers/cm$^2$ without the field and for $H_{ex}=5.8$ Oe. This field is higher than the circumferential anisotropy field in wires ($H_K \approx 3$ Oe). The frequency dispersion has a resonance character which becomes more pronounced with increasing $\bar{p}$ and when no field is applied. The minimum of $S_{21}(f)$ corresponds to the first antenna resonance in wires. The resonance frequency is seen to depend slightly on the values of the concentration and field indicating that the resonance condition for an individual wire is held in composites as well. This is in agreement with the model of the scale-dependent effective medium[3] and also will be proven here for the considered composite configuration by a rigorous analytical method, described in Appendix. Using the value for resonance, the permittivity of the matrix can be estimated: $\varepsilon = (c/(2l f_{res,1}))^2$, which gives the value of about 2.25 for $l = 5$ cm and $f_{res,1} \approx 2$ GHz. The effect of the field is very large near the resonance: for $\bar{p} = 0.35$ centers/cm$^2$ the transmission increases from $-10$ dB up to $-3.5$ dB at $f = 2.1$ GHz, and for $\bar{p} = 0.46$ centers/cm$^2$ it increases from $-27$ dB up to $-12$ dB at $f = 1.7$ GHz. A very large sensitivity of $S_{21}(f)$ near the antenna resonance agrees with the theoretical predictions. At the resonance, the wire impedance strongly affects the current distribution in wires changing both the resonance frequency and absorption. In magnetic wires the surface impedance has a minimum when the magnetization is along the circumference which is realized for a circumferential anisotropy and $H_{ex} = 0$.

The lower surface impedance is the larger are the scattering properties of wires as the embedded antennas. Thus, a deep minimum in transmission is seen when no field is applied. The surface impedance in the wires increases in the presence of the field since the magnetization rotates towards the wire axis. The field equal to the anisotropy field will saturate the magnetization along the axis and the impedance increases as $\sqrt{\tilde{\mu}}$. In this case the transmission spectrum is very broad with the resonance minimum shifted to higher frequencies. This will be explained in detail in the next section on the basis of the effective permittivity concept providing quantitative comparison of the experimental and theoretical results.

Next we investigated the stress effect on the transmission spectra of wire composites. In this case we have to use as inclusions wires with helical anisotropy which show stress-sensitive magnetoimpedance (see Fig. 1). The transmission spectra of such composite are shown in Fig. 6.

Since the wires had also the length of 5 cm the transmission minimum related to the antenna resonance in wires is seen at a similar frequency of 2 GHz. After the sample was loaded and the stress was applied to the whole composite the transmission minimum deepens changing from −12 dB down to −17.5 dB. The application of tensile stress via the matrix strain to wires with a negative magnetostriction establishes the circumferential anisotropy, therefore, lowers the wire surface impedance. As a result, the scattering at the resonance increases and the rate of transmission is low. Whilst the sample was under stress, a relatively large external field is applied to overcome the effect of applied stress and saturate the magnetization along the axis. Estimating the value of stress imposed on wires to be in the range of GPa, this field will be sufficient to saturate the wire magnetization along the axis, thus, realizing the magnetic state with the maximum surface impedance. In this case transmission becomes about −7 dB. Therefore, we have demonstrated that rotating the magnetization in the wires from a circumference to the axis which increases the surface impedance in approximately $\sqrt{\tilde{\mu}}$ times, changes substantially the transmission spectra: from that having a narrow and deep resonance minimum to those with broader dispersion and increased transmission.

It should be noticed that the transmission at minimum seen in Fig. 6 is greater than that in Fig. 5, however the planar concentration is larger. In this case we used wires with smaller diameter so the volume concentration is smaller. Therefore, there are in general a number of critical parameters to control the microwave response such as the wire geometry including its length and diameter, the wire concentration in the composite, the matrix thickness and its permittivity. We believe that optimizing the system it would be possible to realize tunable microwave selective surfaces.

**V. Theoretical scattering spectra and comparison with experiment**

The transmission experiments will now be analyzed quantitatively on the basis of the effective permittivity concept as described in Section II. We will demonstrate that this simple approach works well. In general, it is not possible to assign the effective parameters to a non-continuous material unless its structural scale is much smaller than the wavelength. In the considered composites the wire length is of the order of the wavelength to realize the condition of the antenna resonance. Yet, the microwave response from such composites can be treated as from continuous media. This was validated by numerical simulations and experiments.[4,6,32,33] In Appendix, we also have described a rigorous analytical method for calculating scattering from a particular wire medium configuration, which supports the effective permittivity concept. Furthermore, we will demonstrate that the experimental transmission spectra are consistent with the frequency dependent form of the permittivity $\varepsilon_{eff}$ comprising the averaged wire polarization induced by the incident wave. We will also demonstrate that the dependency of $\varepsilon_{eff}$ on the magnetic properties of the wires predicted by theory quantitatively describes the change in transmission seen experimentally under the effect of the external magnetic field or stress.

The concept of the effective permittivity of wire based composite, which is sensitive to magnetic properties of the inclusions was developed in Ref. 11. Here we calculate $\varepsilon_{eff}$ using Eq. (3) with structural and magnetic parameters relevant to the experiment. It should be noted that for the experimental configuration (predominantly aligned wires) $\varepsilon_{eff}$ will be of a diagonal tensor form with only one component different from the matrix permittivity. The magnetic anisotropy in wires is taken as circumferential and the controlling parameter is an axial magnetic field which will saturate the magnetization along the wire axis, thus ensuring the necessary magnetization change for low and high surface impedances. The permittivity spectra are shown in Fig. 7(a). It is seen that the plots for $H_{ex} = 0$ (circumferential magnetization) have a resonance behavior with a resonance frequency of 2 GHz for $l = 5$ cm, a large absorption peak, and a negative real part of the permittivity in a wide frequency band past the resonance. In the presence of the magnetic field larger than the anisotropy field which saturates the magnetization along the axis and increases the wire surface impedance (see Eqs. (12)-(14)), the dispersion of $\varepsilon_{eff}$ becomes of a relaxation type with a gradual decrease in the real part which is always positive and a very broad losses. At the resonance frequency the dielectric loss parameter decreases more than 6 times in the presence of the field.

We now compute the transmission T and reflection R coefficients for waves incident normally to a slab of thickness $b = 2$ mm of continuous material with the permittivity function shown in Fig. 7(a). Their frequency plots for $H_{ex} = 0$ and $H_{ex} > H_K$ are given in Fig. 7(b). The comparison of the theoretical transmission spectra and the experimental ones seen in Figs. 5 and 6 shows a very good agreement for both magnetic states in the wire. In the case of the circumferential magnetization which exist in wires with a circumferential anisotropy ($H_{ex} = 0$) or in wires with a helical anisotropy under tensile stress the transmission is very low at the level of $-20$ dB for certain wire concentration in the vicinity of the resonance. Both high losses and negative values of the real part of the permittivity strongly oppose the wave propagation. Establishing the magnetization along the wire axis by applying a sufficiently high magnetic field changes the effective permittivity and recovers the wave propagation. The transmission coefficient becomes in the range of $-8$ dB. For fixed values of the wire length and the permittivity of matrix, precise values of T will depend on the wire concentration and its diameter $2a$. The theoretical results are given for $2a = 15$ μm and $p = 0.0022\%$ which correspond to the experimental data of Fig. 4. The agreement in this case is surprisingly very good considering the complexity of the material and many parameters involved.

It should be noticed that the reflection spectra does not show very pronounced resonance behavior and they are much less sensitive to the magnetic structure of the wire. The reason is the existence of absorption. A similar tendency was observed in the experiment. It seems to be a great deficiency for a number of practical applications including stress monitoring of large civil structures for which reflection parameter would be measured. However, changing the structural parameters of the composites such as the matrix permittivity $\varepsilon$, the wire diameter $2a$ and its length $l$, it is possible to realize a more pronounced dispersion of the reflection coefficient R with a high sensitivity to the magnetization direction in wires. This will require larger values of $\varepsilon$ and $2a$, and also smaller $l$ to keep resonance at similar frequency. An example of such behavior is shown in Fig. 8.

The considered composites will also exhibit the reversal of phase in the reflected wave from $\pi$ to $-\pi$ at a frequency near the peak of the imaginary part of the effective permittivity. Since the position of this peak shifts towards higher frequencies when the magnetization in wires is rotated along its axis (by applying a magnetic field higher than the anisotropy field), the phase reversal also will occur at higher frequencies, as shown in Fig. 9(a). This frequency shift could be conveniently used in realistic applications such as stress monitoring, where the transformation of the dispersion curves will be similar but due to the tensile stress. A similar behavior of the reflective coefficient phase was recently reported in high impedance surfaces with mechanically reconfigurable resonance elements.[34] The phase effect is much greater in thicker composite samples allowing multiple reflection. Figure 9(b) demonstrates the dispersion of the reflective coefficient phase for a composite sample of 5 cm thick and having $\varepsilon = 7$ which is typical of dry concrete. The volume concentration of the wire inclusions is taken larger ($p = 0.01\%$) since in this case a high absorption would not be a problem. The phase has frequency oscillations because of multiple reflections which become not possible at the onset of the dispersion in the effective permittivity characterized by an increase in the dielectric loss (10 dB attenuation criterion). In the presence of a magnetic field large enough to saturate the wire magnetization along the axis the dispersion of $\varepsilon_{eff}$ broadens and the frequency oscillations stop at a much lower frequency. For the considered parameters, the frequency shift is more than 1 GHz.

We have demonstrated the magnetic tunability of the phase parameter on the example of a composite sample having magnetic wires with a circumferential anisotropy. Consequently, a controlling parameter was a magnetic field. In composites having wires with a helical anisotropy tuning could be realized by applying a tensile stress. Therefore this method could be used for sensitive remote stress measurements analyzing the frequency trace of the reflected wave phase.

## VI. Conclusion

In this work, we have demonstrated tunable microwave composites containing short CoFeCrSiB wires through the direct free space measurements and effective medium analysis. The transmission spectrum of the proposed wire medium may have a sufficiently narrow bandgap associated with the antenna resonance in wires. A striking feature of this behavior is that changing the magnetic structure in wires by applying a weak magnetic field or stress influence the antenna resonance and opens bandpass. The achieved tunability typically exceeds 10 dB for external fields of few Oe. The reflection spectrum also alters depending on the magnetization in wires. A particular interest for practical application seems to be a considerable shift in the frequency of the phase reversal of the reflected wave. We have briefly discussed a number of applications predicting that the magnetic wire composites are likely to have a large impact on technology of remote sensing and control. Certainly, further intensive investigations are needed to fully characterize the potential of the proposed microwave materials.

We also have demonstrated that it is appropriate to treat the wire composite medium as a homogeneous material with frequency-dispersive effective permittivity. This reduction in complexity is essential to further understanding of microwave properties of magnetic wire composites including prediction of new effects. One example is the case of composites with wires having a non zero off-diagonal component of the surface impedance. It should be possible to excite the wires with a magnetic field of the incident wave, thus, realizing the "optical" activity typical of chiral structures.[35]

**Acknowledgements**

Authors would like to thank ELIRI S.A. and MFTI Ltd (Chisinau, Republic of Moldova) for the technical assistance provided with the U.S. CRDF grant MR2-1024-CH-03.

**Appendix**

In this Section we will demonstrate on the basis of a rigorous statistical analysis of scattering that the electromagnetic response from composite materials filled with wire-inclusions can be described in terms of the effective permittivity. For the sake of simplicity, we will consider a thin composite sheet with predominately oriented wires. We also will be restricted to the case of a normal-incidence plain wave which corresponds to a majority of the experimental data available. On the other hand, this requirement may be of a particular importance since all the wires regardless their length are excited with a uniform electric field. However, any further complications would be beyond the scope of this paper and the complete analysis will be published elsewhere.

The induced current density distribution $j(x)$ in an individual wire taking into account of resistive, radiative and magnetic losses was obtained in Ref. 11:

$$\frac{\partial^2}{\partial x^2}(G*j) + k^2(G*j) = \frac{i\omega\varepsilon}{4\pi}\bar{e}_{0x}(x) - \frac{i\omega\varepsilon\varsigma_{xx}}{2\pi\,ac}(G_\varphi * j), \tag{A1}$$

Here $x$ is the coordinate along the wire, $k = (\omega/c)\sqrt{\varepsilon\mu}$ is the wave number in the matrix, $\varsigma_{xx}$ is the longitudinal component of the surface impedance tensor, and $\bar{e}_{0x}(x)$ is the magnitude of the incident electric field taken at the wire surface. In the case of a normal-incidence plain wave, considered in this Appendix, $\bar{e}_{0x}(x) = const$. The convolutions with the current are of the form:

$$(G*j) = \int_{-l/2}^{l/2} G(x-s)j(s)ds, \tag{A2}$$

$$(G_\varphi * j) = \int_{-l/2}^{l/2} G_\varphi(x-s)j(s)ds. \tag{A3}$$

Here $l$ is the wire length and $G(r)$ is the outgoing Green function of the Helmholtz equation:

$$G(r) = \frac{\exp(ikr)}{4\pi\,r}, \tag{A4}$$

where $r = \sqrt{(x-s)^2 + a^2}$ and $a$ is the wire radius. The function $G_\varphi(r)$ is related with the circular magnetic field $\bar{h}_\varphi(x,a)$ induced by the current at the wire surface:

$$\bar{h}_\varphi(x,a) = \frac{2}{ac}(G_\varphi * j) = \frac{2}{ac}\int_{-l/2}^{l/2} G_\varphi(r)j(s)ds, \tag{A5}$$

$$G_\varphi(r) = \frac{a^2(1-ikr)\exp(ikr)}{2r^3}.$$

Equation (A1) is completed by imposing the boundary conditions $j(-l/2) = j(l/2) \equiv 0$.

In order to take into account the radiative interaction between the wires, the external field $\bar{e}_{0x}$ in Eq. (A1) should be replaced with the total field $\bar{E}_{0x}(x)$ which includes along with $\bar{e}_{0x} = const$ the $x$-components of all scattered fields from the other wires taken at a given wire. Further, we will introduce local coordinate systems for each wire. The electromagnetic field scattered from an individual wire and observed in an arbitrary point $\mathbf{R}(R_x, R_y, R_z)$ has the following form in the wire local coordinate system:

$$\mathbf{e}(R_x, R_y, R_z) = -\frac{4\pi}{i\omega\varepsilon}\left(\left(\frac{\partial^2}{\partial x^2} + k^2\right)(G*j), \frac{\partial^2(G*j)}{\partial x \partial y}, \frac{\partial^2(G*j)}{\partial x \partial z}\right), \qquad (A6)$$

where convolution of $j(x)$ with the Green function $G(r)$ is taken in the point $\mathbf{R}(R_x, R_y, R_z)$ meaning that $r = \sqrt{(R_x - s)^2 + R_y^2 + R_z^2}$.

We are now in a position to formulate Eq. (A1) for the $m$-th wire with the total longitudinal electrical field $\bar{E}_{0x}(x)$ on its surface. Since all the wires are parallel, the total field $\bar{E}_{0x}(x)$ taken on the surface of the $n$-th wire in its local coordinate system will contain only the $x$-components of all scattered fields $\mathbf{e}_{m\neq n}(\mathbf{R})$ and also the initial field $\bar{e}_{0x} = const$:

$$\left(\frac{\partial^2}{\partial x^2} + k^2\right)(G*j_m) = \frac{i\omega\varepsilon}{4\pi}\bar{e}_{0x} - \frac{i\omega\varepsilon\varsigma_{xx}}{2\pi ac}(G_\varphi * j_m) - \left(\frac{\partial^2}{\partial x^2} + k^2\right)\sum_{n\neq m}^{N}(G*j_n)_m, \qquad (A7)$$

where $N$ is the total number of wires in the composite sample. The variable $x$ is the same in each local system. The convolution $(G*j_n)_m$ with the $n$-th wire current is taken in the local system of the $m$-th wire:

$$(G*j_n)_m = \frac{1}{4\pi}\int_{-l/2}^{l/2}\frac{\exp\left(ik\sqrt{(x-s+R_{xm}-R_{xn})^2 + (R_{ym}-R_{yn})^2 + z^2}\right)}{\sqrt{(x-s+R_{xm}-R_{xn})^2 + (R_{ym}-R_{yn})^2 + z^2}}j_n(s)ds, \qquad (A8)$$

where z is a parameter.

Let us now consider the composite in a general coordinate system related with the whole sample and introduce a distribution function $\Psi(\mathbf{R}_1, \mathbf{R}_2,..., \mathbf{R}_N)$ which depends on the radius-vectors $(\mathbf{R}_1, \mathbf{R}_2,..., \mathbf{R}_N)$ pointed at the origins of the local coordinate systems (the center of wire length). It has a natural normalization:

$$\frac{1}{S^N} \int_S (N) \int_S \Psi(\mathbf{R}_1, \mathbf{R}_2,..., \mathbf{R}_N) d\mathbf{R}_1 d\mathbf{R}_2 ... d\mathbf{R}_N \equiv 1, \tag{A9}$$

where $N$-fold integration is taken over sample area $S$. We will now rewrite Eq. (A7) for each wire in the general coordinate system and will consider their sum.

$$\left(\frac{\partial^2}{\partial x^2} + k^2\right) \sum_{m=1}^{N} (G * j_m) = N \frac{i\omega\varepsilon}{4\pi} \bar{e}_{0x} - \frac{i\omega\varepsilon\varsigma_{xx}}{2\pi\, ac} \sum_{m=1}^{N} (G_\varphi * j_m) - \\ - \left(\frac{\partial^2}{\partial x^2} + k^2\right) \sum_{m=1}^{N} \sum_{n\neq m}^{N} (G * j_n)_m \tag{A10}$$

The next step is to average (A10) using the distribution function $\Psi$ satisfying (A9). This produces a self-consistent equation for the averaged current density $j(x)$:

$$\left(\frac{\partial^2}{\partial x^2} + k^2\right)(G * j) = \frac{i\omega\varepsilon}{4\pi} \bar{e}_{0x} - \frac{i\omega\varepsilon\varsigma_{xx}}{2\pi\, ac}(G_\varphi * j) - \left(\frac{\partial^2}{\partial x^2} + k^2\right) \frac{(N-1)}{S} (\tilde{G} * j). \tag{A11}$$

In (A11) the averaged Green function $\tilde{G}$ is calculated with the reduced two-particle distribution function $\tilde{\Psi}_2(\mathbf{R}_1, \mathbf{R}_2)$ (depending on the radius-vectors of any two wires):

$$\tilde{G} = \frac{1}{4\pi\, S} \int_S \int_S \frac{\exp\left(k\sqrt{(x_m - s + R_{x1} - R_{x2})^2 + (R_{y1} - R_{y2})^2 + z^2}\right)}{\sqrt{(x_m - s + R_{x1} - R_{x2})^2 + (R_{y1} - R_{y2})^2 + z^2}} \tilde{\Psi}_2(\mathbf{R}_1, \mathbf{R}_2) d\mathbf{R}_1 d\mathbf{R}_2, \tag{A12}$$

$$\tilde{\Psi}_2(\mathbf{R}_1, \mathbf{R}_2) = \int_S (N-2) \int_S \Psi(\mathbf{R}_1, \mathbf{R}_2,..., \mathbf{R}_N) d\mathbf{R}_3 ... d\mathbf{R}_N,$$

$$\frac{1}{S^2}\iint_{S\,S}\widetilde{\Psi}_2(\mathbf{R}_1,\mathbf{R}_2)d\mathbf{R}_1 d\mathbf{R}_2 \equiv 1.$$

In (A12) the averaging is done with respect to in-plane vectors $\mathbf{R}_1(R_{x1},R_{y1})$ and $\mathbf{R}_2(R_{x2},R_{y2})$, and the parameter $z$ may be taken to be 0 (because the pair correlation will eliminate the singularity in the integral). Assuming the composite is uniform in the plane and taking the statistical limit $\lim_{N\to\infty;\,S\to\infty}(N/S)=\bar{p}$, the equation for the average current density becomes:

$$\frac{\partial^2}{\partial x^2}\left[(G*j)+\bar{p}(\widetilde{G}*j)\right]+k^2\left[(G*j)+\bar{p}(\widetilde{G}*j)\right]=\frac{i\omega\varepsilon}{4\pi}\bar{e}_{0x}-\frac{i\omega\varepsilon\varsigma_{xx}}{2\pi\,ac}(G_\varphi*j), \tag{A13}$$

where $\bar{p}$ is regarded as the in-plane concentration of the wire-centers projected onto the sample surface (this is a number of centers within a given area), and $\widetilde{G}$ is the limit of (A12) with $S\to\infty$.

Equation (A13) should be compared with (A1) for an individual non-interacting wire. One immediate conclusion which can be drawn from this comparison is that there is no renormalization of the wave number $k$ since the left part of (A13) can be written with respect to the same differential operator $(\partial^2/\partial x^2+k^2)$.[5] Therefore, there is no effect of the wire interactions on the antenna resonance frequencies. However, the amplitude of $j(x)$ will depend on the concentration $\bar{p}$. Eventually, the resonance frequency will show some dependence on $\bar{p}$ as a result of the field penetration inside the wire and impedance boundary conditions which is described by the second term in the right part of (A13). This is in agreement with our experimental results (see Fig. 5). Such behavior of antenna resonances in a system of interacting wires was predicted in Refs. 3 and 5 on the basis of scale-dependent effective medium model. However, the situation may be different in the composite with randomly oriented wires. This case requires further investigation.

Next, we will use the equation for the averaged current density to deduce the effective polarization induced by the wire inclusions. We will start with introducing the current density $\mathbf{J}_n(\mathbf{R},t)$ of the $n$-th wire written in the general coordinate system:

$$\mathbf{J}_n(\mathbf{R},t) = \boldsymbol{\tau} \int_{-l/2}^{l/2} j_n(s,t)\delta(\mathbf{R}_n + \boldsymbol{\tau}s - \mathbf{R})ds, \quad (A14)$$

where $\boldsymbol{\tau}$ is a unit vector along the wire inclusions (all are parallel), $\delta(\mathbf{x})$ is the Dirac function with a vector variable $\mathbf{x}$ (the set function), and $\mathbf{R}_n$ is the vector pointed at the center of the $n$-th wire. In a general coordinate system, the current density $\mathbf{J}_n(\mathbf{R},t)$ and the charge density $\rho_n(\mathbf{R},t)$ are related by the continuity equation:

$$\frac{\partial \rho_n(\mathbf{R},t)}{\partial t} + \operatorname{div}\mathbf{J}_n(\mathbf{R},t) = 0, \quad (A15)$$

$$\rho_n(\mathbf{R},t) = \int_{-l/2}^{l/2} g_n(s,t)\delta(\mathbf{R}_n + \boldsymbol{\tau}s - \mathbf{R})ds,$$

where $g_n(x,t)$ is the charge density in the local coordinate system of the $n$-th wire, and $\partial g_n(x,t)/\partial t + \partial j_n(x,t)/\partial x = 0$ is the local continuity equation for the $n$-th wire. Taking time dependence as $\exp(-i\omega t)$ gives:

$$\rho_n(\mathbf{R}) = -\frac{i}{\omega}\operatorname{div}\mathbf{J}_n(\mathbf{R}). \quad (A16)$$

Using (A16), we will write the microscopic Maxwell equation for the total scattered field $\mathbf{e}$:

$$\operatorname{div}\mathbf{e} = 4\pi \sum_{n=1}^{N} \rho_n = -\frac{4\pi i}{\omega} \sum_{n=1}^{N} \operatorname{div}\mathbf{J}_n. \quad (A17)$$

For averaging the scattered field $\mathbf{e}$, it is assumed that the wire inclusions are distributed not only in the plane of composite sheet, but also over its thickness, which is still much smaller comparing to the wavelength and was not essential when finding the average current. In this case the planar concentration should be replaced with the volume concentration according to $\bar{p} = p_V b$, where $b$ should be regarded as the effective thickness. The field **e** is averaged with the volume distribution function $\Phi(\mathbf{R}_1, \mathbf{R}_2, ..., \mathbf{R}_N)$, satisfying the normalization condition:

$$\frac{1}{V^N}\int_V (N) \int_V \Phi(\mathbf{R}_1, \mathbf{R}_2, ..., \mathbf{R}_N) d\mathbf{R}_1 d\mathbf{R}_2 ... d\mathbf{R}_N \equiv 1, \qquad (A18)$$

where $V$ is the volume of composite sample. The averaged scattered field $\mathbf{E} = <\mathbf{e}>$ is of the form:

$$\text{div}\,\mathbf{E} = -\frac{4\pi i}{\omega}\frac{N}{V}\text{div}\left(\boldsymbol{\tau}\int_V \int_{-l/2}^{l/2} j_1(s)\delta(\mathbf{R}_1 + \boldsymbol{\tau} s - \mathbf{R})\tilde{\Phi}_1(\mathbf{R}_1)dVds\right) = \\ = -\frac{4\pi i}{\omega}\frac{N}{V}\text{div}\left(\boldsymbol{\tau}\int_{-l/2}^{l/2} j_1(s)\tilde{\Phi}_1(\mathbf{R} - \boldsymbol{\tau} s)ds\right) \qquad (A19)$$

where $\tilde{\Phi}_1(\mathbf{R}_1) = \frac{1}{V^{N-1}}\int_V (N-1)\int_V \Phi(\mathbf{R}_1, \mathbf{R}_2, ..., \mathbf{R}_N) d\mathbf{R}_2 ... d\mathbf{R}_N$ is the reduced single-particle distribution function. Provided the composite is uniform ($\tilde{\Phi}_1 \equiv 1$), and considering $\lim_{N\to\infty; V\to\infty}(N/V) = p_V$, Eq. (A19) allows us to deduce the polarization of this medium ($\text{div}\,\mathbf{E} = -4\pi\,\text{div}\,\mathbf{P}$):

$$\mathbf{P} = \boldsymbol{\tau}\frac{i}{\omega}p_V \int_{-l/2}^{l/2} j(s)ds = \boldsymbol{\tau}\, p_V \mathscr{P} = \boldsymbol{\tau}\frac{\bar{p}}{b}\mathscr{P}, \qquad (A20)$$

where $j(x)$ is the average current defined by (A11) and (A12) with $\bar{e}_{0x} = const$, and $\mathscr{P} = \frac{i}{\omega}\int_{-l/2}^{l/2} j(s)ds \equiv \int_{-l/2}^{l/2} g(s)s\,ds$ is the dipole moment of an individual wire induced by the average current density $j(x)$ with the condition $j(-l/2) = j(l/2) \equiv 0$. Thus, it has been demonstrated that the composite with the wire inclusions can be characterized by the local effective permittivity $\varepsilon_{\mathit{eff}\|}$:

$$\varepsilon_{\mathit{eff}\|} = \varepsilon + 4\pi\, p_V \mathscr{P} = \varepsilon + 4\pi\, (\bar{p}/b)\mathscr{P} \tag{A21}$$

when the electrical field in the normal-incidence plane wave is directed along wires. For the transverse polarization, for which $\mathscr{P} \equiv 0$, we obtain $\varepsilon_{\mathit{eff}\perp} = \varepsilon$. Therefore, the effective permittivity of a thin wire composite sample takes the diagonal tensor form.

We have to admit that taking $\lim\limits_{N\to\infty;\,V\to\infty}(N/V) = p_V$ and averaging in (A19) are not strictly correct since the effective thickness $b$ is in fact an unknown parameter (it is not a physical thickness of the composite). Furthermore, $\lim\limits_{N\to\infty;\,V\to\infty}(N/V) = p_V$ requires remoteness of the boundaries, which certainly does not take place in the case of an optically thin composite sheet. Nevertheless, the use of the effective permittivity appears to be correct when solving external problems of finding scattered fields in the radiative near or far field region. However, it should be realized that $b$ is some free parameter not necessarily coinciding with the composite thickness.

**References**

[1] D. F. Sievenpiper, E. Yablonovitch, J. N. Winn, S. Fan, P. R. Villeneuve, and J. D. Joannopoulos, Phys. Rev. Lett. **80**, 2829 (1998).

[2] J. B. Pendry, A. J. Holden, W. J. Stewart, and I. Youngs, Phys. Rev. Lett. **76**, 4773 (1996).

[3] A. N. Lagarkov and A. K. Sarychev, Phys. Rev. B **53**, 6318 (1996).

[4] A. N. Lagarkov, S. M. Matitsine, K. N. Rozanov, and A. K. Sarychev, J. App. Phys. **84**, 3806 (1998).

[35]L. V. Panina, D. P. Makhnovskiy, and K. Mohri, J. Magn. Magn. Mat. **272-276**, 1452 (2004).

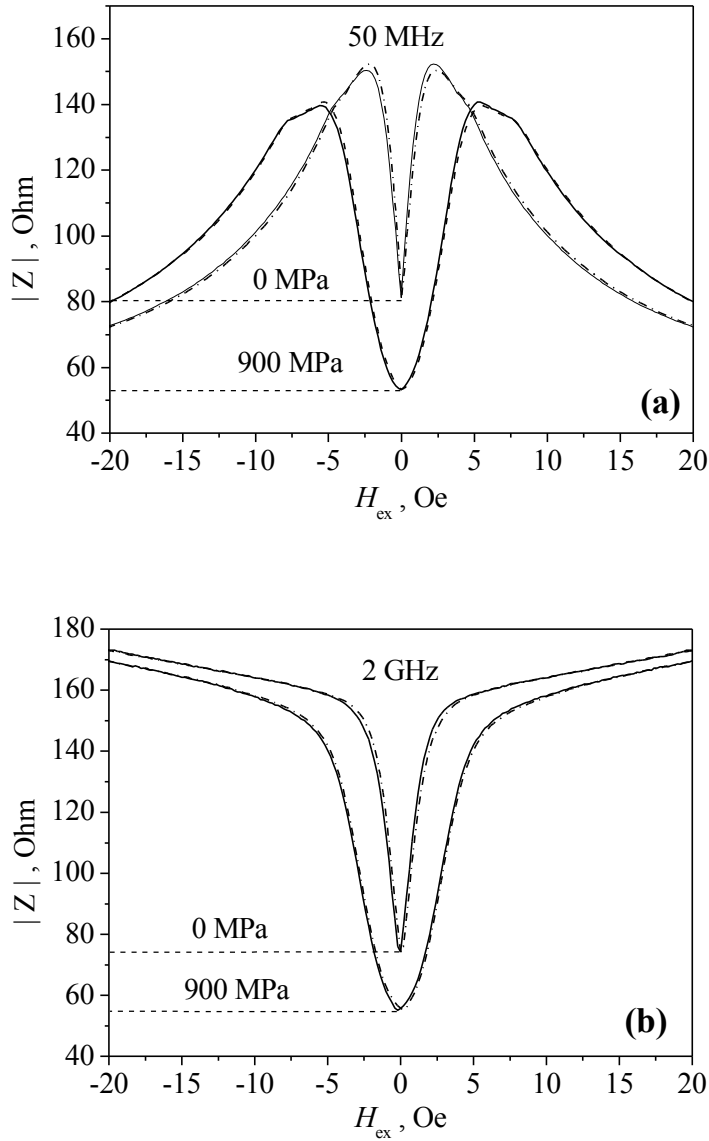

**Fig. 1.** Impedance plots vs. external field measured in CoFeCrSiB wires with a helical anisotropy and external tensile stress as a parameter ($\sigma_{ex} = 0$ and $\sigma_{ex} = 900$ MPa). Frequency is 50 MHz in (a) and 2 GHz in (b). As the frequency is increased the impedance peaks seen in (a) tend to broaden and at GHz frequencies the impedance shows very gently sloping for high fields $H_{ex} > H_K$. Applying tensile stress establishes a circumferential anisotropy and the impedance value drops at $H_{ex} = 0$.

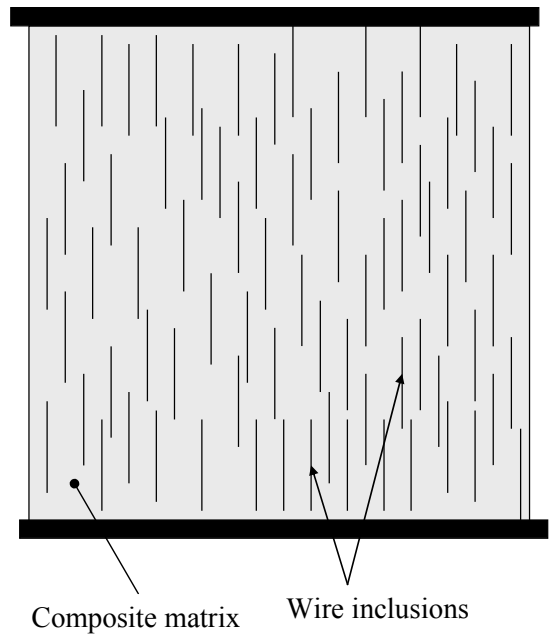

Fig. 2. Sample of composite materials with predominantly aligned wire inclusions.

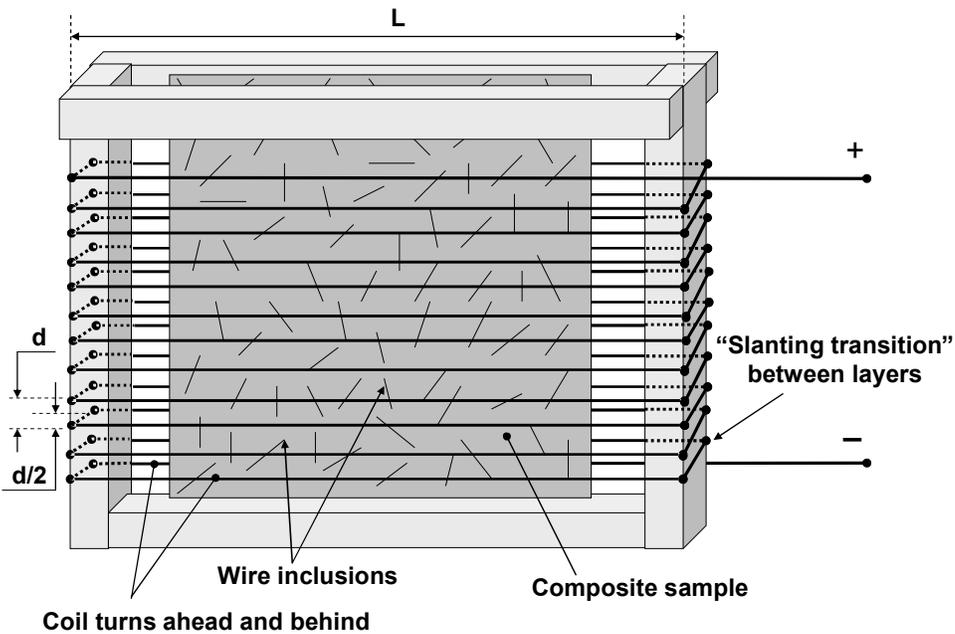

**Fig. 3.** Schematic view of the planar coil generating the uniform magnetic field. The distance between the coil layers $d = 1$ cm is equal to the distance between the turns, and the turns in two layers are shifted by $d/2$. The sample is placed in the middle of the coil. The coil becomes "invisible" for a plane-polarised wave with the electrical vector directed transversely to the coil turns.

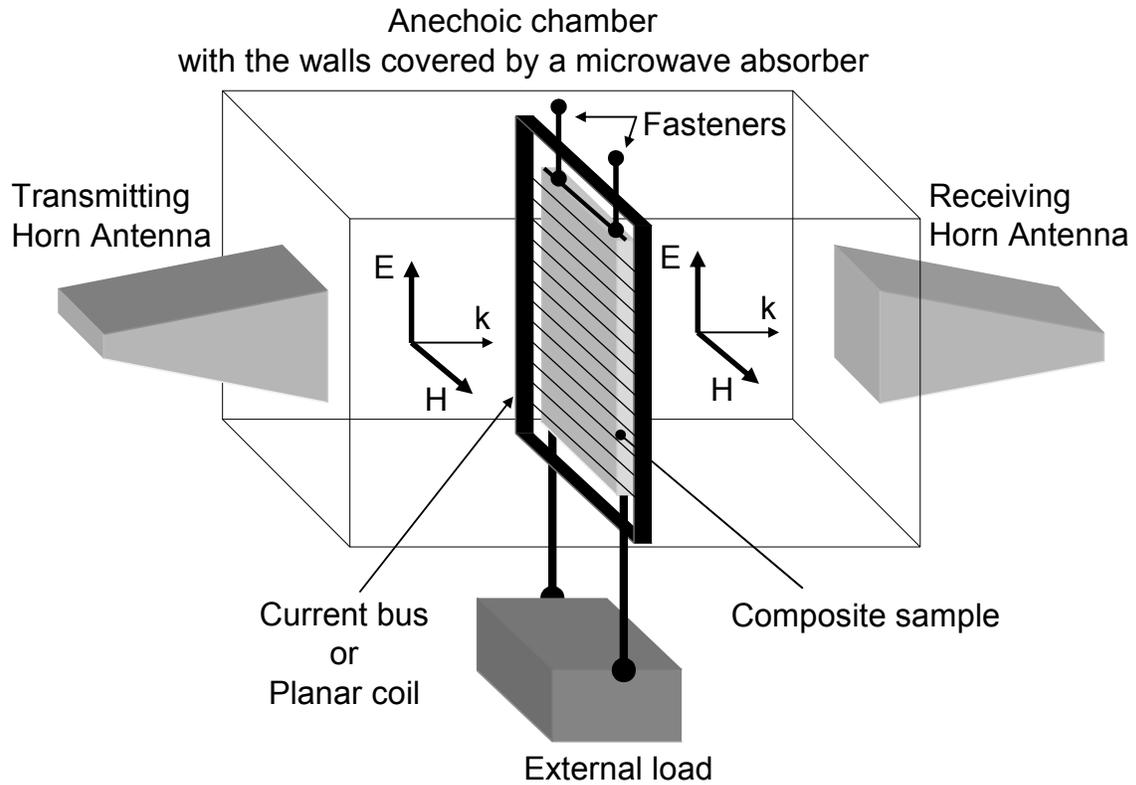

**Fig. 4.** Schematic view of the experimental free space set up comprising broad band horn antennas, a compact anechoic chamber with a composite sample placed in the planar coil. External load can be also applied to the composite sample.

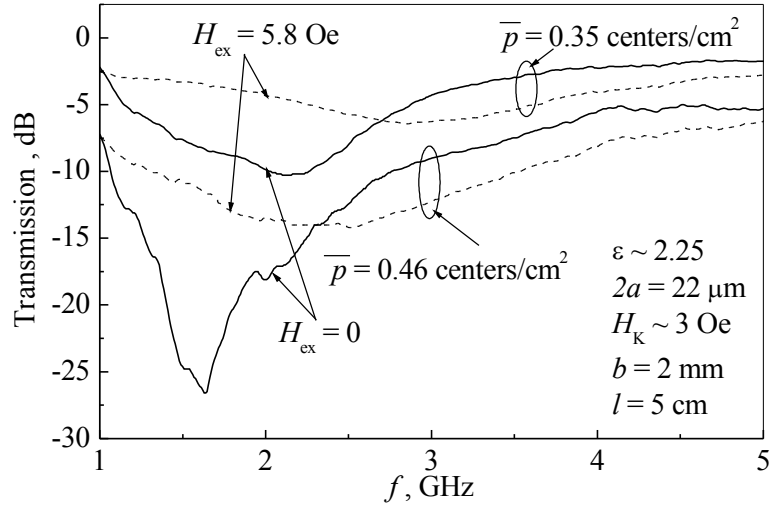

**Fig. 5.** Transmission spectra measured in the composite containing CoFeCrSiB amorphous wires with a circumferential anisotropy for two values of the in-plane concentration $\bar{p} = 0.35$ and $\bar{p} = 0.46$ centers/cm$^2$. The data are given for $H_{ex} = 0$ and $H_{ex} = 5.8$ Oe. A transmission minimum seen at $H_{ex} = 0$ has a tendency to tail up in the presence of a sufficiently large field $H_{ex} > H_K$.

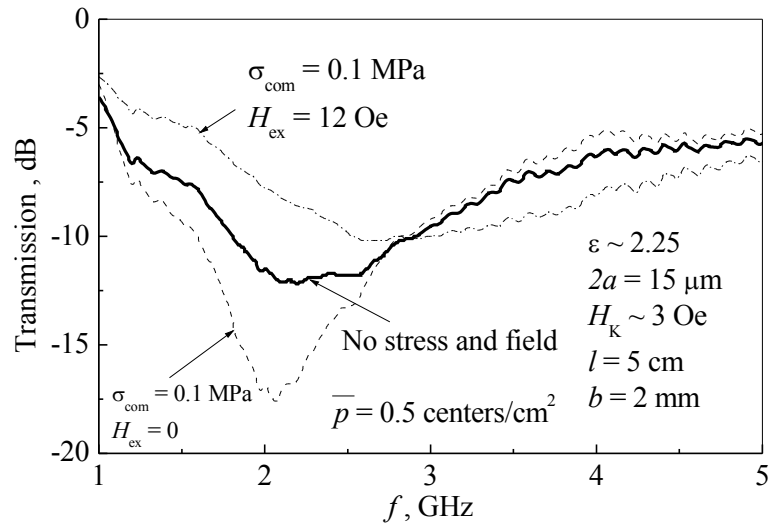

**Fig. 6.** Transmission spectra measured in the composite material containing CoFeCrSiB amorphous wires having a helical anisotropy with the external stress and field as parameters. The plot when no external stimuli are applied is between those with $\sigma_{ex}$ and ($\sigma_{ex}$, $H_{ex}$) as the dc magnetisation in wires changes from helical to circumferential and axial, respectively.

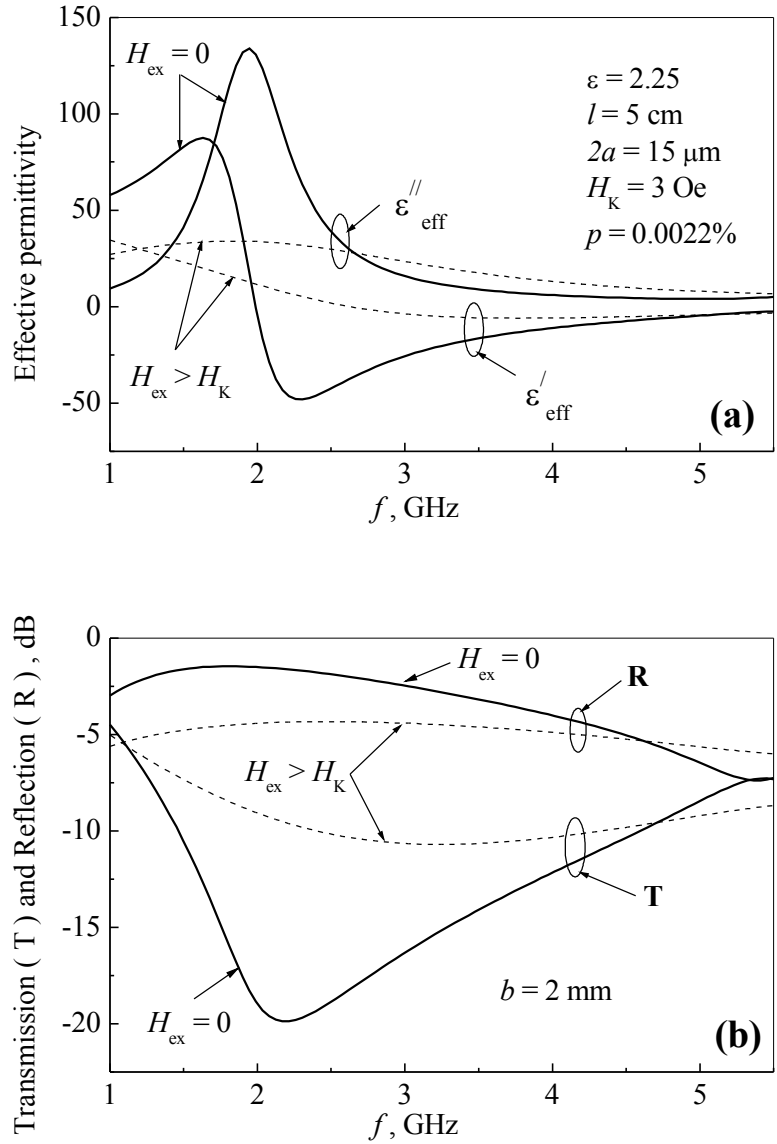

**Fig. 7.** Theoretical dispersion of the effective permittivity in (a) and coefficients of transmission and reflection in (b) for a composite containing wires with a circumferential anisotropy. The plots are shown for two limiting cases: $H_{ex} = 0$ (circumferential magnetisation and low impedance) and $H_{ex} > H_K$ (axial magnetisation and high impedance). Notice that there exists a frequency band with a negative real part of $\varepsilon_{eff}$ for $H_{ex} = 0$ where transmission is strongly suppressed.

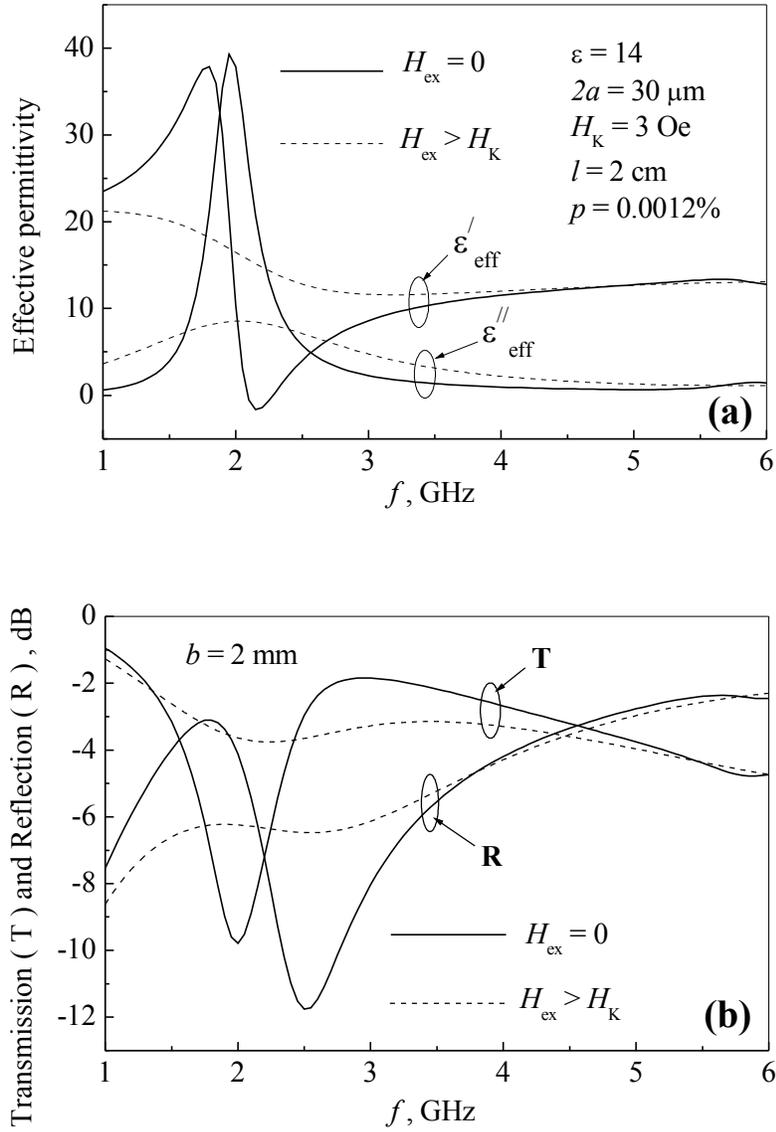

**Fig. 8.** Theoretical dispersion of the effective permittivity in (a) and coefficients of transmission and reflection in (b) for a composite with high permittivity matrix $\varepsilon = 14$ containing thicker wires with a metal core diameter of 30 microns. The plots are shown for two limiting cases of $H_{ex} = 0$ and $H_{ex} > H_K$. Both reflection and transmission parameters show a resonance dispersion and high sensitivity to the magnetic state of the wires.

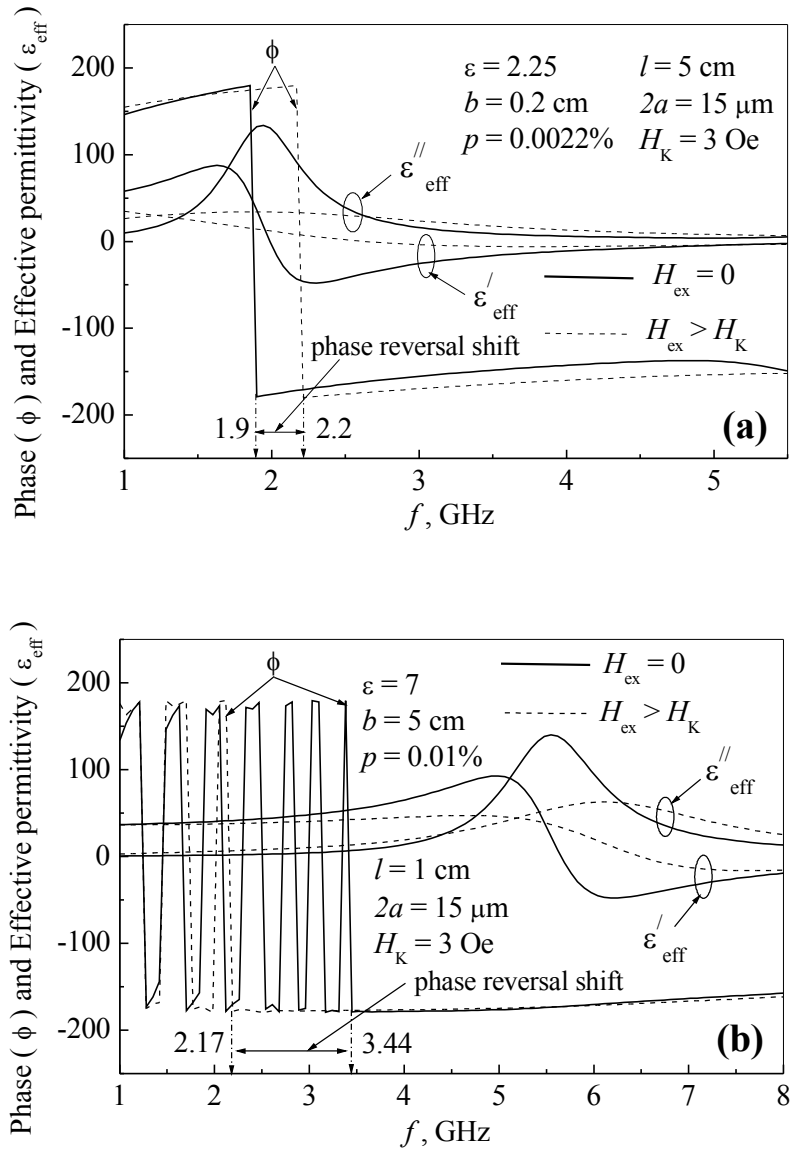

**Fig. 9.** Theoretical dispersions of the phase of reflection coefficient for thin (a) and thick (b) composites, respectively, with the external field as a parameter. For comparison, the permittivity dispersion is given as well.